\documentclass[pre,twocolumn]{revtex4}
\usepackage{epsfig}
\usepackage{bm}
\usepackage{amsmath}
\usepackage{hyperref}
\usepackage{breakurl}

\begin{document}

\title{Helical velocity fluctuations and large-scale circulation in turbulent thermal convection}

\author{A. Bershadskii}

\affiliation{
ICAR, P.O. Box 31155, Jerusalem 91000, Israel
}

\begin{abstract}

 It  is shown that helicity dynamics dominates spectral distribution  of the velocity fluctuations in turbulent thermal convection at moderate and large values of the Rayleigh number (distributed chaos and scaling respectively). The large-scale circulation (wind) in the turbulent Rayleigh-B\'{e}nard convection is especially sensitive to the helicity dynamics and its chaotic reversals can be associated with the reversals of the mean helicity sign (the ergodic restoration of the reflectional symmetry in a long run). Results of the laboratory experiments and measurements in the free atmospheric convection have been used in order to support the theoretical considerations based on the adiabatic invariance of the second order moment of the helicity distribution. The geomagnetic field dipole variability and reversals have been also discussed in this context and good agreement with results of numerical simulations and observational data has been established.
\end{abstract}

\maketitle

\section{Inroduction}

  Dynamics of velocity and temperature fluctuations in the turbulent thermal convection is an old and difficult problem (see for reviews Refs. \cite{lx},\cite{ch},\cite{vkp}). The pioneering works of Bolgiano and Obukhov \cite{ob},\cite{bol} applied the Kolmogorov-like phenomenology to stably stratified fluids and then their results were extended on the the Rayleigh-B\'{e}nard convection in the Refs. \cite{pz},\cite{lvov},\cite{fl}. The scaling power spectra theoretically predicted in these papers for the velocity fluctuations are rarely observed in the laboratory experiments and atmospheric measurements. It will be shown in present paper that the main reason for this is helical nature of the velocity fluctuations in the thermal convection. 
  
  It will be shown (Section II) that although helicity is not conserved  in the thermal convection (even in the inviscid approximation) the second order moment of the helicity distribution (the Levich-Tsinober invariant of the Euler equation \cite{lt},\cite{mt}) is an inviscid invariant of the thermal convection at certain (rather general) conditions. Application of the Kolmogorov-like phenomenology using this adiabatic invariant results in the scaling power spectrum $E(k) \propto k^{-4/3}$ for the velocity fluctuations in the inertial range of scales (Section III). Usually the scaling inertial range is not large (if it appears at all) for the velocity fluctuations in the turbulent thermal convection and the distributed chaos approach is often providing a more adequate description of the processes (especially for the moderate values of the Rayleigh number). Corresponding to this adiabatic invariant power spectrum $E(k) \propto \exp-(k/k_{\beta})^{1/3}$  will be obtained for the velocity fluctuations in the frames of the distributed chaos approach (Section IV). \\
  
  A large-scale circulation (clockwise or anticlockwise global circular wind) can appear in the turbulent Rayleigh-B\'{e}nard convection at sufficiently large Rayleigh numbers (see, for instance, Refs, \cite{niemela},\cite{sbn} and references therein). Appearance of such global circulation in a cylindrical cell, for instance, violates the reflectional spatial symmetry. Natural attempts of the system to restore the reflecional symmetry in a long run result in the abrupt chaotic \cite{b} reversals of the direction of the large-scale circulation. It is shown (Section V) that the large-scale circulation is especially sensitive to the helicity dynamics and the reversals of the direction of the circulation can be associated with corresponding reversals of the mean helicity's sign. \\
  
  It is believed that the geomagnetic field is generated by a thermal convection-driven geomagnetic dynamo (see for a review Ref. \cite{amit}). The temporal variability and chaotic abrupt reversals of the geomagnetic dipole has been discussed in the Section VI in the above mentioned context and good agreement with results of numerical simulations of the thermal convection-driven geomagnetic dynamo and observational data has been established.

\section{Adiabatic invariants}

  In the Boussinesq approximation thermal (buoyancy driven) convection is described by equations \cite{kcv}
$$
\frac{\partial {\bf u}}{\partial t} + ({\bf u} \cdot \nabla) {\bf u}  =  -\frac{\nabla p}{\rho_0} + \sigma g \theta {\bf e}_z + \nu \nabla^2 {\bf u}   \eqno{(1)}
$$
$$
\frac{\partial \theta}{\partial t} + ({\bf u} \cdot \nabla) \theta  =  S  \frac{\Delta}{H}e_z u_z + \kappa \nabla^2 \theta, \eqno{(2)}
$$
$$
\nabla \cdot \bf u =  0 \eqno{(3)}
$$
where $\theta$ is the temperature fluctuations (over the temperature profile), ${\bf u}$ is the velocity and $p$ is the pressure, ${\bf e}_z$ is a unit vector (along the gravity direction) and $g$ is the gravity acceleration, $H$ and $\Delta$ are the distance between the layers and the temperature difference between the layers, the mean density is denoted as $\rho_0$, whereas $\nu$, $\kappa$ and $\sigma$ are the viscosity, thermal diffusivity and thermal expansion coefficient.  For the unstable stratification (Rayleigh-B\'{e}nard convection) $S=+1$ whereas for the stable stratification $S=-1$.\\   

  In the non-dissipative approximation ($\nu=\kappa=0$) equations (1-3) have a generalized energy invariant 
$$
\mathcal{E} = \int_V ({\bf u}^2 -S\sigma g \frac{H}{\Delta}\theta^2) ~ d{\bf r}   \eqno{(4)}
$$    
where $V$ is the spatial domain's volume \cite{kcv}).\\

  At certain conditions the Eqs (1-3) have an additional inviscid invariant. For $\nu =0$ equation for mean helicity is 
$$
\frac{d\langle h \rangle}{dt}  = 2\sigma g e_z\langle \omega_z \theta \rangle \eqno{(5)} 
$$ 
(the helicity density - $h={\bf u}\cdot {\boldsymbol \omega}$, the vorticity - ${\boldsymbol \omega} = \nabla \times {\bf u}$, $\langle...\rangle$ - average over the spatial volume $V$), hence the helicity is not an inviscid invariant of the thermal convection. Let us recall that the mean helicity together with the mean energy are the fundamental invariants for the Euler equations \cite{mt}. The Euler equations have an additional invariant - the second order moment of the helicity distribution (the Levich-Tsinober invariant \cite{lt},\cite{mt}). In the thermal convection the main contribution to the correlation $\langle \omega_z \theta \rangle$ (from the Eq. (5)) comes from the large-scale motion (mainly from the coherent structures) and it can be rather considerable, but the correlation between $\omega_z$ and $\theta$ is quickly diminished with decreasing spatial scales due to the turbulent effects. In order to take into account this phenomenon let us consider the second order moment of the helicity distribution. To define this moment let us divide the spatial domain of motion into the cells (with volumes $V_j$ and the boundary conditions ${\boldsymbol \omega} \cdot {\bf n}=0$ on the surfaces of the cells) moving with the fluid \cite{mt}. The second order moment can be then defined as \cite{mt}
$$
I = \lim_{V \rightarrow  \infty} \frac{1}{V} \sum_j H_{j}^2  \eqno{(6)}
$$
where 
$$
H_j = \int_{V_j} h({\bf r},t) ~ d{\bf r}.  \eqno{(7)}
$$
 Due to above mentioned phenomenon the $H_j$ are inviscid quasi-invariants for the cells with small enough spatial scales. For sufficiently developed turbulence such cells provide the main contribution to the sum Eq. (6) (cf. Ref. \cite{bt}) and, as a consequence, the total sum in the Eq. (6) is also a quasi-invariant at $\nu = 0$.
 
\begin{figure} \vspace{-2cm}\centering
\epsfig{width=.52\textwidth,file=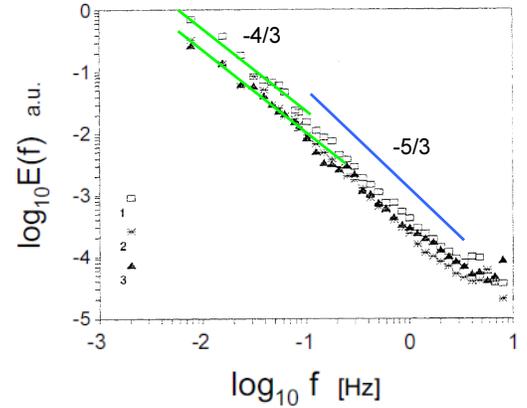} \vspace{-6cm}
\caption{Power spectrum of the longitudinal velocity fluctuations in the atmospheric near free convection.} 
\end{figure}

\section{Scaling spectra of the velocity fluctuations}

  In the frames of the Kolmogorov-Obukhov phenomenology \cite{my} scaling spectra of the velocity fluctuations in the inertial range of scales can be estimated as 
$$
E(k) \propto \varepsilon^{2/3} k^{-5/3},   \eqno{(8)}
$$
where  
$$  
\varepsilon = \left|\frac{d\langle{\bf u}^2 -S\sigma g \frac{d}{\Delta}\theta^2 \rangle}{dt}\right| \eqno{(9)}
$$
and the generalized energy Eq. (4) is considered as an adiabatic invariant in the inertial range of scales. \\

  In order to apply this approach using the second order moment of the distribution of the helicity density - $I$ (Eq. (6)), as an adiabatic invariant in the inertial range of scales we should take into account that (unlike the energy, which is a quadratic invariant) the second order moment $I$ is a quartic invariant (cf. Eq. (6)). Therefore, we should use $\varepsilon_I =|dI^{1/2}/dt|$ instead of $\varepsilon$, and we obtain from the dimensional considerations
$$
E(k) \propto \varepsilon_I^{2/3} k^{-4/3}  \eqno{(10)}
$$
  
 Results of atmospheric measurements of the longitudinal velocity fluctuations in the near free convection (Rayleigh-B\'{e}nard convection, i.e. the unstable stratification conditions) were reported in the Ref. \cite{gr}. The measurements were made by a probe at the height $\sim$ 12m above sea surface. The surface was aerodynamically smooth and the weather was calm. The results for three data sets with the wind gusts comparable to the r.m.s. horizontal velocity fluctuations (near free convection) are shown in the log-log scales in Figure 1 (the spectral data for the Fig. 1 have been taken from Fig. 1a of the Ref. \cite{gr}). The solid straight lines are drawn to indicate the scaling spectra: Eq. (8) for the small-scale part of the inertial range and Eq. (10) for the large-scale part of the inertial range (the "frozen-in" Taylor hypothesis \cite{gr},\cite{kv} were applied in order to compare the frequency spectra shown in the Fig. 1 with the wavenumber spectra Eqs. (8) and (10)). \\

\section{Distributed chaos} 

 Let us start from a simple (Lorenz-) model of the thermal convection considering only the first modes in a Galerkin approximation of the Eqs. (1-3) \cite{lorenz}
$$
\frac{dx}{dt} = \sigma (y - x),~~      
\frac{dy}{dt} = r x - y - x z, ~~
\frac{dz}{dt} = x y - b z   \eqno{(11)}             
$$
 In this system the variables $x(t)$, $y(t)$ and $z(t)$ are proportional to the effective rate of convection and to the horizontal and vertical temperature variations respectively \cite{spa}.\\

  For certain values of the parameters $\sigma~,r,$ and $b$ the deterministic chaotic dynamics was discovered in this system \cite{lorenz},\cite{spa}. As we will see later on in this paper the Lorenz system can be still rather inspiring. \\

  Figure 2 shows a typical example of Lorenz chaotic (strange) attractor. The trajectory passes around the two fixed point of attraction of the focus type and chaotically switches from one to another forming the two wings of the attractor. Figure 3 shows a short sample of the $X(t)$ in the chaotic regime. The invariance of the Eq. (11) under the (reflective) coordinate transformation 
$$  
  ( x , ~y , ~z ) ~ \longrightarrow ~( -x,~ -y,~ z)   \eqno{(12)}
$$  
(called rotation symmetry around the z-axis) will be useful for understanding of the helical nature of the abrupt chaotic inversions (reversals) of the large-scale circulation (wind) in the real thermal convection (see next Section). \\

  The Lorenz system is a bounded and smooth dynamical system. Deterministic chaos in such systems is usually associated with the exponential frequency spectrum \cite{oh}-\cite{fm}
$$ 
E(f) \propto \exp(-f/f_c)  \eqno{(13)}
$$
 
 Figure 4, for instance, shows in the semi-log scales power spectrum for the $X(t)$-component of the Lorenz system. The dashed straight line indicates the exponential spectrum Eq. (13) and the short dotted arrow indicates position of the $f_c$.\\

 In the spatial domain the frequency spectrum Eq. (13) corresponds to the wavenumber spectrum 
$$ 
E(k) \propto \exp(-k/k_c)  \eqno{(14)}
$$ 
(see Ref. \cite{mm} and references therein).\\

\begin{figure} \vspace{-1.3cm}\centering
\epsfig{width=.49\textwidth,file=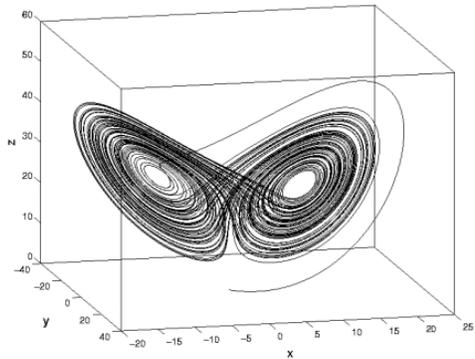} \vspace{-5.6cm}
\caption{A typical example of Lorenz chaotic attractor.} 
\end{figure}
\begin{figure} \vspace{-0.3cm}\centering
\epsfig{width=.45\textwidth,file=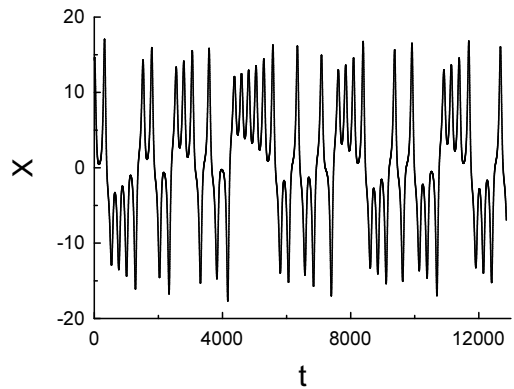} \vspace{-4.6cm}
\caption{A sample of the $X(t)$ time series for the Lorenz system.} 
\end{figure}
\begin{figure} \vspace{-0.9cm}\centering
\epsfig{width=.48\textwidth,file=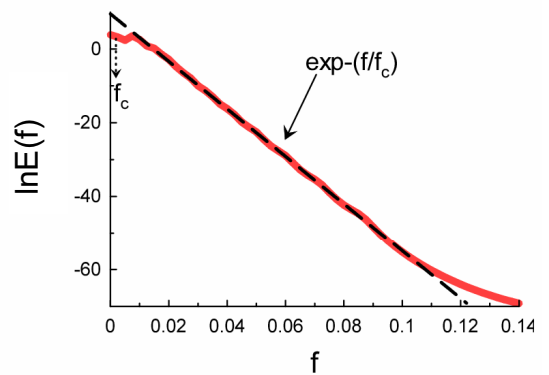} \vspace{-4.7cm}
\caption{Power spectrum of the $X(t)$ for the Lorenz system.} 
\end{figure}

  Increase of the number of the modes under consideration to a more realistic value results in fluctuations of the parameter $f_c$ in Eq. (13) and parameter $k_c$ in the Eq. (14).

 An ensemble averaging over the fluctuating exponential spectrum
$$
E(k) \propto \int_0^{\infty} P(k_c) \exp -(k/k_c)dk_c  \eqno{(15)}
$$ 
can account this phenomenon and the stretched exponential spectrum 
$$
E(k) \propto \exp-(k/k_{\beta})^{\beta} \eqno{(16)}
$$ 
can be considered as a generalization of the simple exponential one.\\

 Comparing the Eqs. (15) and (16) we obtain an estimation of the large $k_c$ asymptotic of the probability distribution $P(k_c)$ in the Eq. (15) \cite{jon}
$$
P(k_c) \propto k_c^{-1 + \beta/[2(1-\beta)]}~\exp(-\gamma k_c^{\beta/(1-\beta)}) \eqno{(17)}
$$     
where $\gamma$ is a constant.\\
\begin{figure} \vspace{-1.15cm}\centering
\epsfig{width=.45\textwidth,file=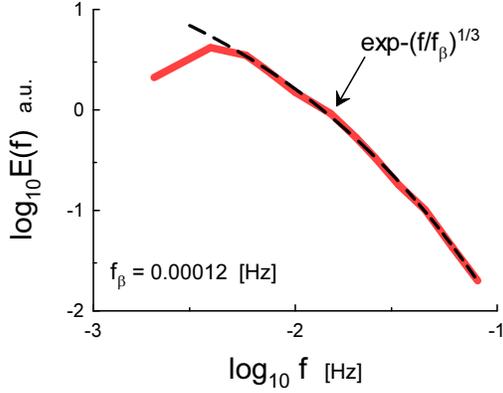} \vspace{-4.4cm}
\caption{Power spectrum of the local vertical velocity at $Ra = 1.2 \cdot 10^8$.} 
\end{figure}
\begin{figure} \vspace{-1cm}\centering
\epsfig{width=.45\textwidth,file=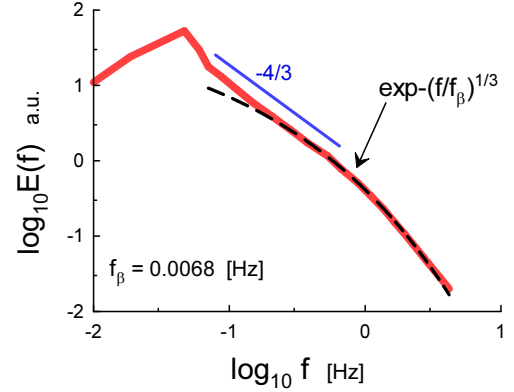} \vspace{-4.6cm}
\caption{Power spectrum of the local vertical velocity at $Ra = 2.3 \cdot 10^{10}$.} 
\end{figure}
\begin{figure} \vspace{-0.5cm}\centering
\epsfig{width=.45\textwidth,file=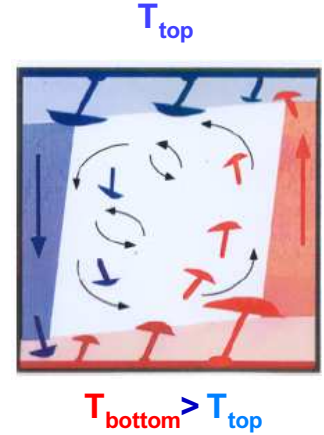} \vspace{-3.95cm}
\caption{Kadanoff's cartoon of generation of the large-scale circulation in the turbulent Rayleigh-B\'{e}nard convection.} 
\end{figure}
  On the other hand, the probability distribution $P(k_c)$ can be obtained using simple dimensional considerations. Indeed, the scaling behaviour of the characteristic velocity $v_c$ for helicity dominated process can be estimated from the dimensional considerations as
$$
v_c \propto I^{1/4} k_c^{1/4}.     \eqno{(18)}
$$
It directly follows from the Eq. (18) that if $v_c$ has a Gaussian distribution  \cite{my}, then
$$
P(k_c) \propto k_c^{-3/4}\exp-bk^{1/2}   \eqno{(19)}
$$
Comparing Eqs. (17) and (19) and taking $\gamma =b$ one obtains $\beta = 1/3$, i.e. 
$$
E(k) \propto \exp-(k/k_{\beta})^{1/3}.   \eqno{(20)}
$$
where $k_{\beta}$ is a renormalized (due to the fluctuations) characteristic wavenumber.\\

In Ref. \cite{xia4} results of the laser Doppler velocimetry measurements of Rayleigh-B\'{e}nard convection in a cylindrical cell (filled with water) were reported. Local vertical velocity near the vertical sidewall of the cell was measured at mid-height of the cell at different values of the Rayleigh number (from $Ra = 1.2 \cdot 10^8$ to $Ra = 2.3 \cdot 10^{10}$). A large-scale circulation was also present in this experiment and the measurements were made within the rotational plane of the circulation. The aspect ratio of the cell $\Gamma=D/H$ (where $D$ is the cell's diameter) was close to 1.\\

  Figures 5 and 6 show power spectra of the local vertical velocity at $Ra = 1.2 \cdot 10^8$ and $Ra = 2.3 \cdot 10^{10}$, respectively (the spectral data were taken from Fig. 2 of the Ref. \cite{xia4}). The dashed curve indicates correspondence to the spectral law Eq. (20) (again the Taylor hypothesis has been applied). The straight line with the slope "-4/3" (cf. Eq. (10)) has been drawn in the Fig. 6 for reference.

\section{Large-scale circulation}

   Figure 7 schematically shows how the cold (blue) and hot (red) plumes generate the large-scale circulation in the turbulent Rayleigh-B\'{e}nard convection (adapted from Ref. \cite{kad}). This schematic picture was confirmed by direct observations in the laboratory experiments (see, for instance, Ref. \cite{xia8} and references therein).\\
\begin{figure} \vspace{-1.5cm}\centering
\epsfig{width=.48\textwidth,file=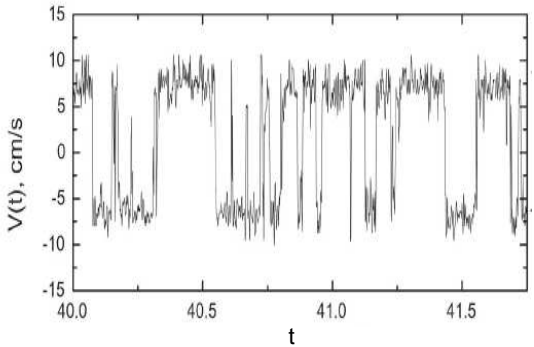} \vspace{-5.45cm}
\caption{ A short sample of the time series of the large-scale circulation velocity at $Ra = 1.5 \cdot 10^{11}$.} 
\end{figure}
\begin{figure} \vspace{-0.45cm}\centering
\epsfig{width=.45\textwidth,file=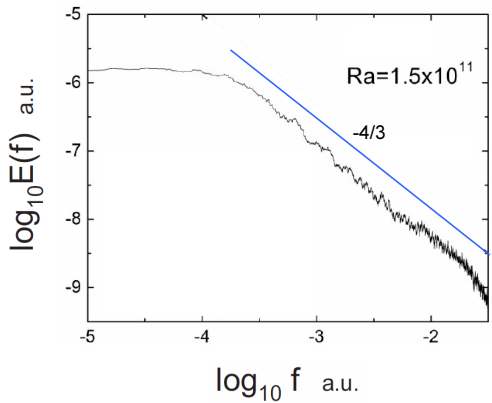} \vspace{-4.37cm}
\caption{Power spectrum of the large-scale circulation velocity at $Ra = 1.5 \cdot 10^{11}$.} 
\end{figure}
\begin{figure} \vspace{-1.7cm}\centering
\epsfig{width=.45\textwidth,file=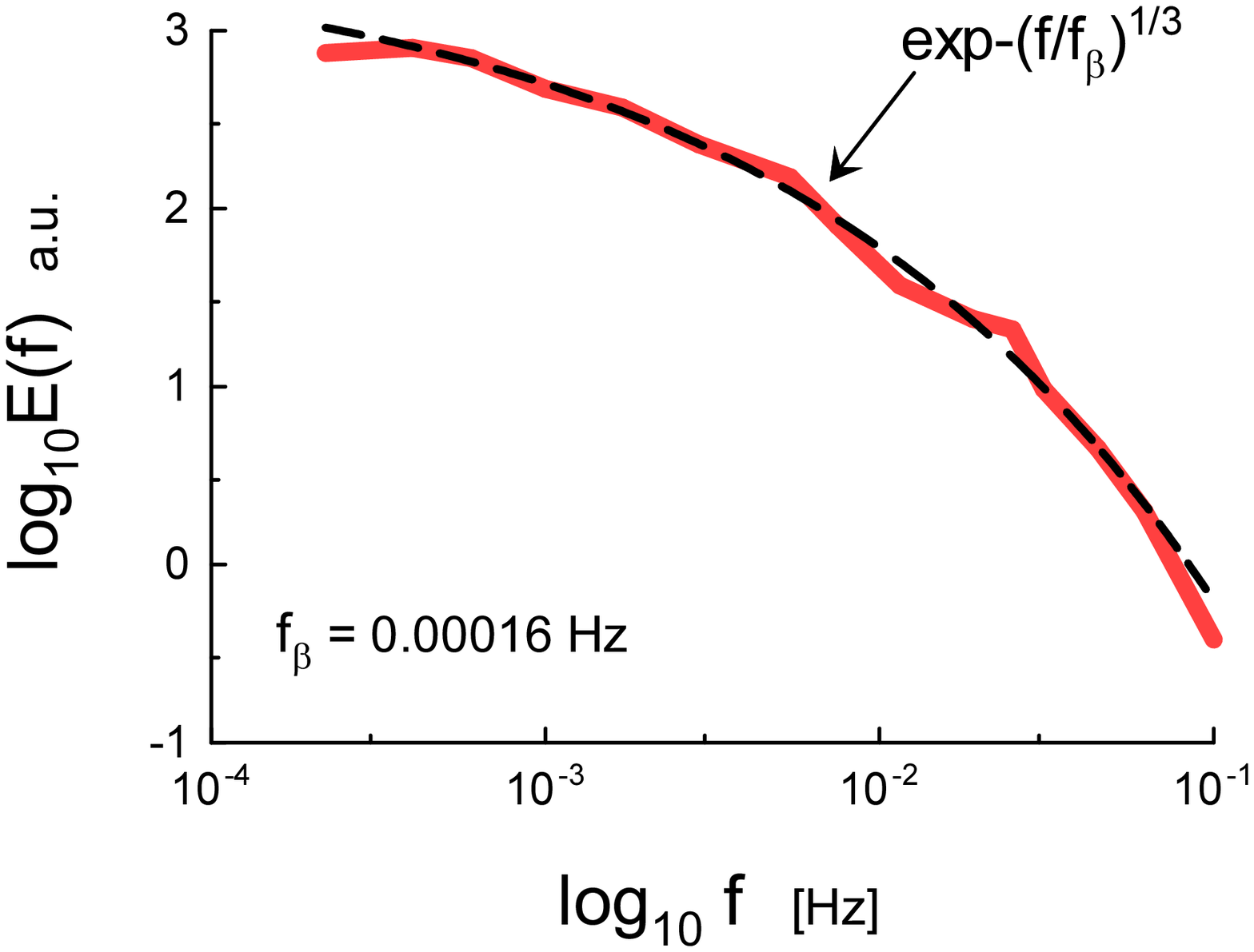} \vspace{-4.1cm}
\caption{Power spectrum of the large-scale circulation velocity at $Ra = 5.6 \cdot 10^{9}$ (at the top of the cell).} 
\end{figure}
\begin{figure} \vspace{-0.55cm}\centering
\epsfig{width=.45\textwidth,file=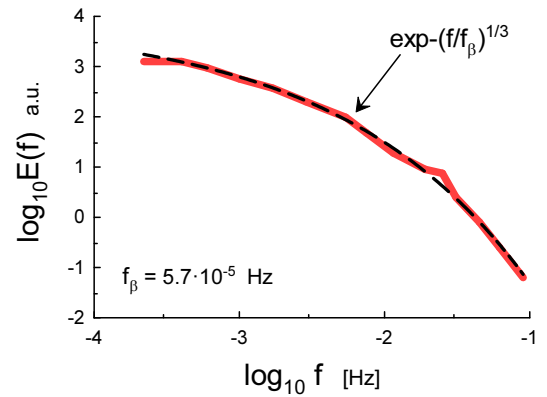} \vspace{-4.33cm}
\caption{Power spectrum of the large-scale circulation velocity at $Ra = 5.6 \cdot 10^{9}$ (at the bottom of the cell).} 
\end{figure}

  Figure 8 (adapted from the Ref. \cite{niemela}) shows a short sample of the time series of the large-scale circulation velocity measured in a cylindrical cell with $\Gamma =1$ (filled with cryogenic helium gas) near the vertical sidewall at mid-height of the cell at rather large value of the Rayleigh number $Ra = 1.5 \cdot 10^{11}$. Figure 9 shows power spectrum of the large-scale circulation velocity (the spectral data were taken from Fig. 2 of the Ref. \cite{niemela}). The straight line is drawn in the Fig. 9 to indicate correspondence to the spectral law Eq. (10). \\
  
  It should be noted that the large-scale circulation was also well observed at considerably smaller values of the Rayleigh number $Ra = 5.6 \cdot 10^{9}$ in a water-filled cylindrical Rayleigh-B\'{e}nard convection cell ($\Gamma =1$) \cite{xia8}, for instance. 
  
    Figures 10 and 11 show the large-scale circulation velocity spectra obtained in this experiment at the top and at the bottom of the cell correspondingly. Figure 12 shows the large-scale circulation velocity spectrum obtained in this experiment at the top of the cell at $Ra = 5.7 \cdot 10^{10}$ and $\Gamma =1/2$ (at the bottom of the cell the spectrum is analogous). The spectral data were taken from Fig. 7 of the Ref. \cite{xia8}.\\

    The dashed curves in the Figs. 10-12 indicate correspondence to the spectral law Eq. (20). One can see that in this case (unlike the previous one - Fig. 9) the spectral behaviour indicates the distributed chaos instead of the scaling. In the both experiments, however, the large scale circulation was dominated by the second order moment of helicity distribution (the Levich-Tsinober invariant) - Eqs. (10) and (18-20).\\ 
\begin{figure} \vspace{-1.5cm}\centering
\epsfig{width=.45\textwidth,file=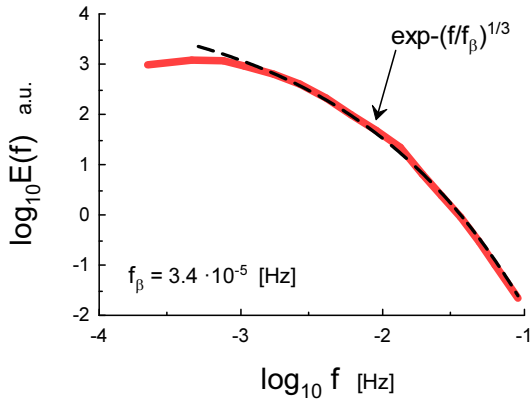} \vspace{-4.35cm}
\caption{Power spectrum of the large-scale circulation velocity at $Ra = 5.7 \cdot 10^{10}$ (at the top of the cell).} 
\end{figure}
      
    The appearance of the clockwise (or alternatively the anticlockwise) large-scale circulation in the cylindrical cell is, of course, a spontaneous breaking of the reflectional symmetry. The observed phenomenon of the chaotic reversals of direction of the large-scale circulation (see Fig. 8, for instance) can be considered as an ergodic restoration of the global reflectional symmetry in a long run (cf. also Fig. 3 and its relation to the reflectional symmetry). Therefore, we can expect that (as in the case of the Lorenz attractor, see Section IV) we have deal with a two-wing attractor and chaotic transitions between the wings resulting in the wind reversals. Then, naturally, the each wing of the attractor can be associated with the mean helicity of the opposite sign (let us recall that the mean helicity is not conserved in the Rayleigh-B\'{e}nard convection) and the chaotic reversals of the large-scale circulation can be associated with reversals of the mean helicity's sign at the chaotic transitions from one wing of the attractor to another (cf. Ref. \cite{l} and references therein).
    
\section{The geomagnetic field dipole variability}

  The global geomagnetic field plays important role in sustaining life on the Earth providing a shield against the energetic charged particles coming from space and the Sun (see, for instance, Refs. \cite{llr},\cite{lam} and references therein). The modern technology, based on the electromagnetic processes, is also effectively shielded from the solar wind effects by the global geomagnetic field. However, the global geomagnetic field intensity is varying in time, and although the characteristic time scales of this variability are usually large (or very large) in comparing with the human lifetime, it seems to be important to understand its temporal dynamics. It should be noted that the current significant decrease of the geomagnetic field intensity is already a problem \cite{hla}. The complete polarity reversals and excursions of the global magnetic field are of especial interest (see, for instance, Ref. \cite{amit} and references therein). \\

   In the Boussinesq approximation the dynamo action induced by thermal convection in an unstably
stratified shell between two concentric spheres rotating with angular velocity $\Omega$ is described by the system of non-dimensional equations
$$
E\cdot\left\{\frac{\partial {\bf u}}{\partial t} + ({\bf u} \cdot \nabla) {\bf u} - \nabla^2 {\bf u}\right\} + \nabla p = {\bf F} ({\bf u},{\bf B}, T) \eqno{(21)}
$$
$$
{\bf F} ({\bf u},{\bf B}, T)  = 2 [\hat{{\bf z}}\times {\bf u}] + Ra\frac{{\bf r}}{R}T + Pm^{-1} [(\nabla \times {\bf B})\times {\bf B}] \eqno{(22)}
$$

$$
\frac{\partial T}{\partial t} + ({\bf u} \cdot \nabla) T  =  Pr^{-1} \nabla^2 T \eqno{(23)}
$$

$$
\frac{\partial {\bf B}}{\partial t} = \nabla \times ( {\bf u} \times
    {\bf B}) +Pm^{-1} \nabla^2 {\bf B}    \eqno{(24)} 
$$

$$
\nabla \cdot {\bf u} =  0, ~~~~~~~~~~ \nabla \cdot {\bf B} =  0 \eqno{(25)}
$$
with appropriate boundary conditions (see, for instance, Ref. \cite{amit} and references therein). In these equations $T$ is temperature, ${\bf B}$ is magnetic field, $\hat{{\bf z}}$ is a unit vector in the direction of the axis of rotation, ${\bf r}$ is the position vector and $R$ is the outer radius of the shell. Dimensionless $E = \nu/ \Omega H^2$ is the Ekman number ($H$ is the shell gap), $Pr = \nu/\kappa$ is the Prandtl number, $Pm = \nu/\eta$ is the magnetic Prandtl number ($\eta$ is the magnetic diffusivity), $Ra = \sigma g \Delta T H/\nu \Omega$ is the modified Rayleigh number ($\Delta T$ is the temperature difference between outer and inner boundaries of the shell). \\

   For inviscid fluid ($\nu=0$) the equation for mean helicity is 
$$
\frac{d\langle h \rangle}{dt}  = 2\langle {\boldsymbol \omega}\cdot {\bf F} ({\bf u},{\bf B}, T) \rangle \eqno{(26)} 
$$ 
(cf. the Eq. (5)). Hence the helicity is not an inviscid invariant in this case as well. However, the second order moment of the helicity distribution $I$ at certain conditions can be an adiabatic invariant of the system Eqs. (21-25). Indeed, 
if the main contribution to the generally non-zero correlation $\langle {\boldsymbol \omega}\cdot {\bf F} ({\bf u},{\bf B}, T) \rangle$ is provided by the large-scale (mainly coherent) fields, but the correlation $\langle {\boldsymbol \omega}\cdot {\bf F} ({\bf u},{\bf B}, T) \rangle$ is quickly weakened toward smaller spatial scales (due to the randomization produced by the turbulent effects), then the consideration similar to that of the Section II supports the adiabatic invariance of the second order moment of the helicity distribution $I$. Therefore, the two types of the spatial (wavenumber) spectrum related to this adiabatic invariance - Eqs. (10) and (20), can be also obtained for the system Eqs. (21-25). In this Section we will be interested in the true temporal (frequency) spectra of the velocity field. The scaling temporal spectrum corresponding to the spatial one Eq. (10) can be readily obtained from the dimensional considerations:
$$
E(f) \propto \varepsilon_I^{4/5} f^{-7/5}   \eqno{(27)}
$$
As for the distributed chaos one should replace the spatial (wavenumber) Eq. (18) by corresponding temporal (frequency) relationship
$$
v_c \propto I^{1/5} f_c^{1/5}.     \eqno{(28)}
$$
Then, in the way similar to that used in the Section IV one obtains for the distributed chaos the stretched exponential temporal (frequency) spectrum
$$
E(f) \propto \exp-(f/f_{\beta})^{2/7}.   \eqno{(29)}
$$
where $f_{\beta}$ is a renormalized (due to the fluctuations) characteristic frequency (cf. the Eq. (20)).\\

  The equations (21-25) are dimensionless, but since the spectra Eqs. (27) and (29) have been obtained from the dimensional considerations let us recall that in the Alfv\'enic units the magnetic field has the same dimension as velocity. Therefore, in this units the same dimensional considerations that were used in order to obtain the velocity spectrum Eq. (27) or Eq. (29) can be also used to obtain spectrum of the magnetic field in the same form of the Eq. (27) or (29) (cf., for instance, Ref. \cite{b2} and references therein). \\
\begin{figure} \vspace{-1.7cm}\centering
\epsfig{width=.45\textwidth,file=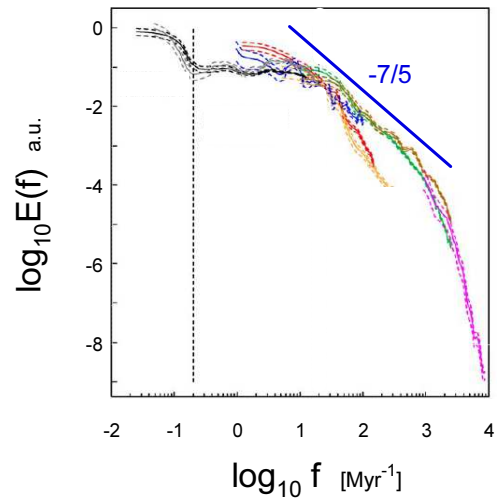} \vspace{-3.2cm}
\caption{Composite power spectrum of geomagnetic dipole moment variations for the time interval 0-160 Myr.} 
\end{figure}
\begin{figure} \vspace{-0.5cm}\centering
\epsfig{width=.45\textwidth,file=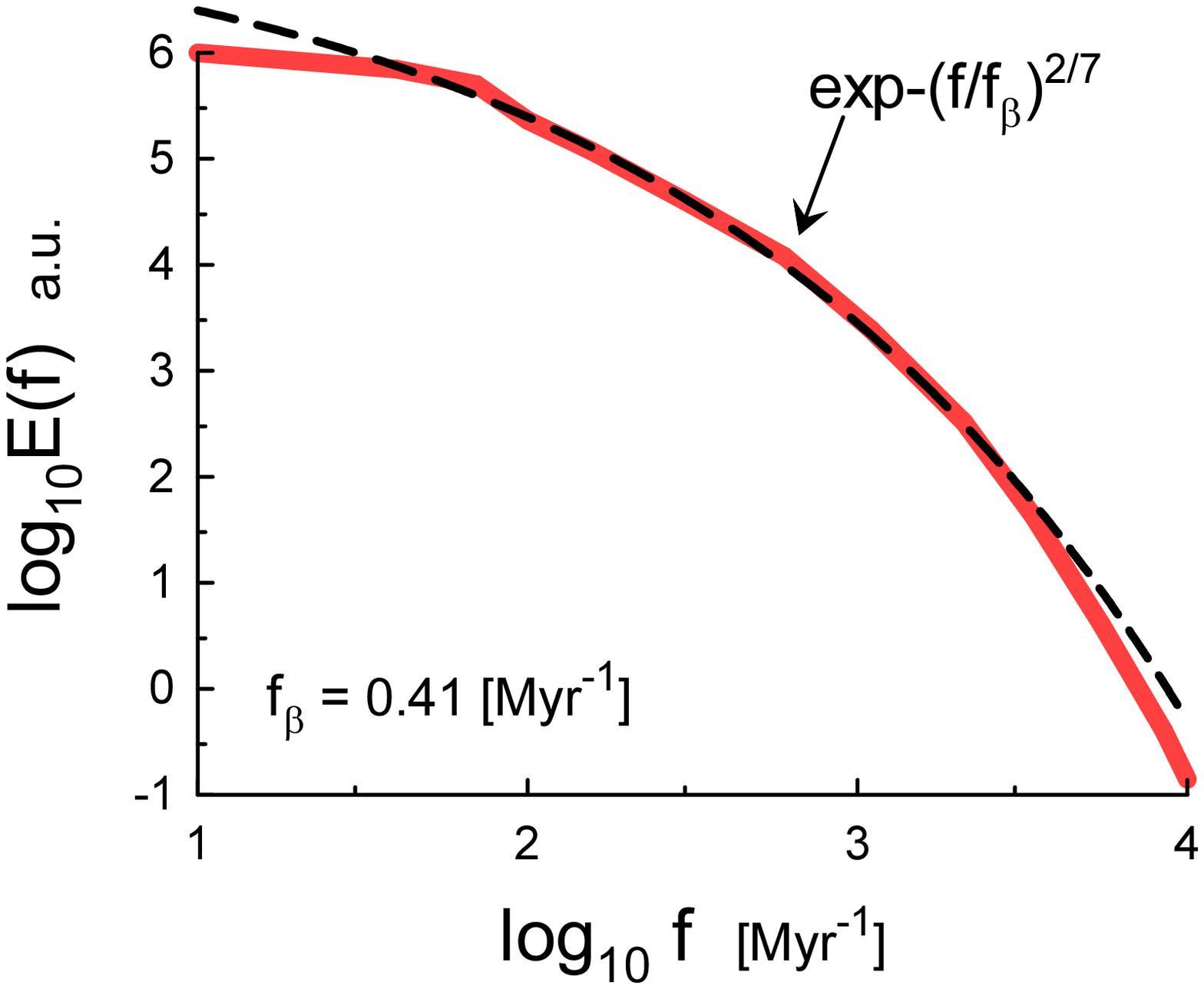} \vspace{-4.45cm}
\caption{Power spectrum of the axial dipole magnetic field (numerical simulation, $Rm = 564$).} 
\end{figure}

   The paleo-geomagnetic data is usually obtained for the geomagnetic dipole moment. The magnetic dipole moment normalized by the domain's volume $V$ is
$$
{\boldsymbol \mu} =   \frac{1}{2V} \int [{\bf r} \times {\bf j}]~ dV = \frac{1}{2V} \int [{\bf r} \times (\nabla \times {\bf B})]~ dV  \eqno{(30)}
$$
and in the Alfv\'enic units the ${\boldsymbol \mu}$ also has the same dimension as velocity, and the same dimensional considerations can be used in this case as well. Therefore the scaling Eq. (27) and stretched exponential Eq. (29) spectra can be used for the geomagnetic dipole moment dynamics.  \\

   Figure 13 shows a composite power spectrum of the geomagnetic field dipole variability and reversals for the time interval 0-160 Myr (the spectrum was adapted from the Fig. 7 of the Ref. \cite{cj}). The magnetostratigraphic time
scale, different marine sediment paleointensity records and a paleomagnetic field model were used to construct the composite spectrum. The solid straight line is drawn to indicate the scaling spectrum Eq. (27).\\

  Figure 14 shows power spectrum of the axial dipole magnetic field obtained in a thermal convection-driven geodynamo simulation reported in Ref. \cite{dc} (the spectral data were taken from Fig. 3 of the Ref \cite{dc} for the magnetic Reynolds number $Rm =564$). The system Eqs. (21-25) were used at this simulation with appropriate boundary conditions and some additional constraints (see for details Ref. \cite{dc}). The dashed curve is drawn to indicate the spectrum Eq. (29) (the distributed chaos). \\
\begin{figure} \vspace{-1.5cm}\centering
\epsfig{width=.45\textwidth,file=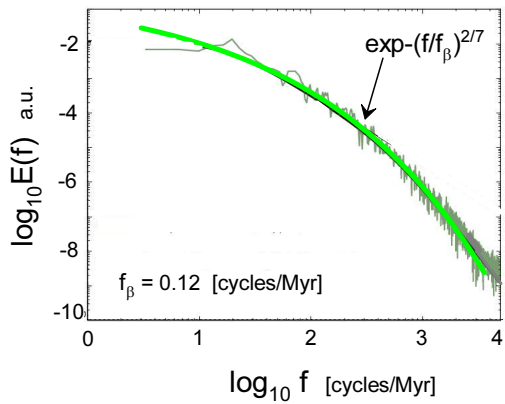} \vspace{-4.5cm}
\caption{Power spectrum of the axial dipole magnetic
 field (numerical simulation, $Rm = 90$).} 
\end{figure}
\begin{figure} \vspace{-0.5cm}\centering \hspace{-1.5cm}
\epsfig{width=.55\textwidth,file=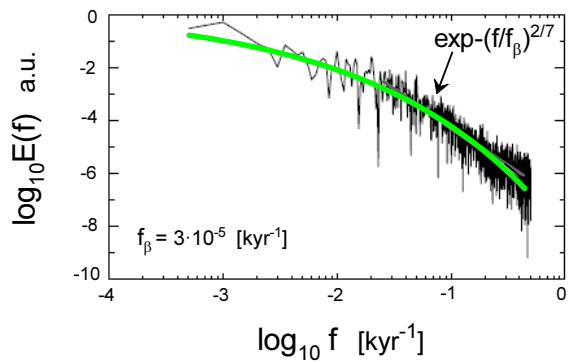} \vspace{-6.65cm}
\caption{Power spectrum of the the virtual axial dipole moment. The SINT-2000 data set for the the past 2 Myr.}
\end{figure}

  Figure 15 shows power spectrum of the axial dipole magnetic field obtained in another thermal convection-driven geodynamo simulation reported in Refs. \cite{bkm},\cite{bm} (the spectral data were taken from Fig. 2 of the Ref \cite{bm} for the magnetic Reynolds number $Rm =90$). The system Eqs. (21-25) were used at this simulation with appropriate boundary conditions at $E = 5 \cdot 10^{−5}$, $Ra = 1400$, $Pm= 0.5$ and $Pr = 1$. The lime curve is drawn to indicate the spectrum Eq. (29).\\
  
  The geomagnetic dipole moment variability for the past two million years were studied in the Ref. \cite{vmg} using a global composite stack of the records of relative magnetic palaeointensity obtained from sediment cores in different oceanic basins (the SINT-2000 data set). It was shown in the Ref. \cite{vmg} that the obtained dipole dynamics is in good agreement with that obtained from the volcanic lavas. It was also noted that the axial dipole rebuilds itself after reversal (in the opposite direction) in a few thousand years, i.e. the reversals are rather abrupt. Figure 16 shows power spectrum of the virtual axial dipole moment variability computed using this data set (the spectral data were taken from Fig. 12b of the Ref. \cite{mori}). The lime curve in the Fig. 16 is drawn to indicate correspondence to the spectrum Eq. (29).\\

  One can conclude that the thermal convection-driven geomagnetic dipole variability is apparently dominated by the second order moment of the helicity distribution - $I$, in an inertial range of scales. Comparing this result with the above discussed abrupt chaotic reversals of the large-scale circulation in the pure thermal convection one can expect that the observed abrupt chaotic reversals of the geomagnetic field polarity are related to the ergodic reversals in mean helicity sign as well, and the observed dynamics of the geomagnetic field can provide information about dynamics of velocity field in the Earth's liquid core.


\begin{thebibliography}{99}
\bibitem{lx} D. Lohse and K-Q. Xia, Annu. Rev. Fluid Mech., {\bf 42}, 335 (2010)
\bibitem{ch} E.S.C. Ching, Statistics and Scaling in Turbulent Rayleigh-B\'{e}nard Convection (Singapore: Springer, 2014)
\bibitem{vkp} M.K. Verma, A. Kumar and A. Pandey, New J. Phys., {\bf 19}, 025012 (2016)
\bibitem{ob} A.M. Obukhov, Dokl. Akad. Nauk SSSR, {\bf 125}, 1246 (1959)
\bibitem{bol} R. Bolgiano, J. Geophys. Res., {\bf 64}, 2226 (1959)
\bibitem{pz} I. Procaccia and R. Zeitak, Phys. Rev. Lett., {\bf 62}, 2128 (1989)
\bibitem{lvov} V.S. L'vov, Phys. Rev. Lett., {\bf 67}, 687 (1991)
\bibitem{fl} G. Falkovich and V.S. L'vov, Physica D, {\bf 57}, 85 (1992)
\bibitem{lt} E. Levich and A. Tsinober, Phys. Lett. A {\bf 93}, 293 (1983)
\bibitem{mt} H.K. Moffatt and A. Tsinober, Annu. Rev. Fluid Mech., {\bf 24}, 281 (1992)
\bibitem{niemela} J.J. Niemela, L. Skrbek, K.R. Sreenivasan and R.J. Donnelly, J. Low Temp. Phys., {\bf 126}, 297 (2002)
\bibitem{sbn} K.R. Sreenivasan, A Bershadskii, J.J. Niemela, Phys. Rev. E, 056306  (2002)
\bibitem{b} A. Bershadskii, Chaos, {\bf 20}, 043124 (2010)
\bibitem{amit} H. Amit, R. Leonhardt and J. Wicht, Space Sci. Rev., {\bf 155}, 293 (2010)
\bibitem{kcv} A. Kumar, A.G. Chatterjee and M.K. Verma, Phys. Rev. E, {\bf 90}, 023016 (2014)
\bibitem{bt} A. Bershadskii and A. Tsinober,  Phys. Rev. E, {\bf 48}, 282 (1993).
\bibitem{my} A. S. Monin, A. M. Yaglom, Statistical Fluid Mechanics, Vol. II: Mechanics of Turbulence (Dover Pub. NY, 2007)
\bibitem{gr} A.A. Grachev, Boundary-Layer Meteorology, {\bf 69}, 27 (1994)
\bibitem{kv} A. Kumar and M.K. Verma, R. Soc. open sci., {\bf 5}, 172152 (2018)
\bibitem{lorenz} E.N. Lorenz, J. Atmos. Sci., {\bf 20}, 130 (1963)
\bibitem{spa} C. Sparrow, The Lorenz Equations: Bifurcations, Chaos, and Strange Attractors (Springer-Verlag, 1982)
\bibitem{oh} N. Ohtomo, K. Tokiwano, Y. Tanaka et. al., J. Phys. Soc.
Jpn., {\bf 64}, 1104 (1995)
\bibitem{sig} D.E. Sigeti, Phys. Rev. E, {\bf 52}, 2443 (1995)
\bibitem{f} J. D. Farmer, Physica D, {\bf 4}, 366 (1982).
\bibitem{fm}U. Frisch and R. Morf, Phys. Rev., {\bf 23}, 2673 (1981)
\bibitem{mm} J. E. Maggs and G. J. Morales, Phys. Rev. Lett., {\bf 107},
185003 (2011); Phys. Rev. E {\bf 86}, 015401(R) (2012); Plasma Phys. Control.
Fusion, {\bf 54}, 124041 (2012)
\bibitem{jon} D.C. Johnston, Phys. Rev. B, {\bf 74}, 184430 (2006)
\bibitem{xia4} X-D. Shang and K-Q. Xia, Phys. Rev E, 64, 065301(R0 (2001)
\bibitem{kad} L.P. Kadanoff,  Annu. Rev. Cond. Matt. Phys., {\bf 6}, 1 (2015)
\bibitem{xia8} H.-D. Xi and K.-Q. Xia, Phys. Rev. E, {\bf 78}, 036326 (2008)
\bibitem{l} E. Levich, Concept. Phys., {\bf VI}(3), 239 (2009)
\bibitem{llr} R. Lundin, H. Lammer, and I. Ribas, Space Sci. Rev.,
{\bf 129}, 245 (2007)
\bibitem{lam} H. Lammer, J.H. Bredehoft, and A. Coustenis, et al., Astron. Astrophys. Rev., {\bf 17}, 181 (2009)
\bibitem{hla} G. Hulot, F. Lhuillier, and J. Aubert, Geophy. Res. Lett., {\bf 37}, L06305( 2010)
\bibitem{b2} A. Bershadskii, arXiv:2002.10984 (2020)
\bibitem{cj} C. Constable and C. Johnson, Phys. Earth and Planet. Interiors, {\bf 153}, 61 (2005)
\bibitem{dc} J. Davies and C.G. Constable, Earth and Planet. Sci. Lett., {\bf 404}, 238 (2014)
\bibitem{bkm} B.A. Buffett, E.M. King and H. Matsui, Geophys. J. Int., {\bf 198}, 597 (2014)
\bibitem{bm} B.A. Buffett and H. Matsui, Earth and Planet. Sci. Lett., {\bf 411}, 20 (2015)
\bibitem{vmg} J-P. Valet, L. Meynadier and Y. Guyodo, Nature, {\bf 435}, 802 (2005)
\bibitem{mori} N. Mori, D. Schmitt, J. Wicht, A. Ferriz-Mas, H. Mouri, A. Nakamichi and M. Morikawa, Phys. Rev. E, {\bf 87}, 012108 (2013)


\end{thebibliography}
\end{document}